\documentclass{article}
\usepackage{srcltx}
\usepackage[english]{babel}
\usepackage{amsmath}
\usepackage{amssymb}
\usepackage{amsfonts}
\usepackage{latexsym}
\usepackage{textcomp}
\usepackage{appendix}
\usepackage{multirow}
\usepackage{booktabs}
\usepackage{rotating}
\usepackage{afterpage}
\usepackage{pdflscape}
\usepackage{url}
\usepackage{array}
\usepackage{marvosym}
\usepackage{graphics}
\usepackage{graphicx}
\usepackage{natbib}
\usepackage[table]{xcolor}
\usepackage{epsfig}
\usepackage{float}
\usepackage{authblk}

\usepackage[caption=false]{subfig}
\newcommand{\overbar}[1]{\mkern 1.5mu\overline{\mkern-1.5mu#1\mkern-1.5mu}\mkern 1.5mu}
\usepackage[margin=1.25in]{geometry}

\title{Are cryptocurrencies becoming more interconnected?}

\author[1]{Nektarios Aslanidis }
\author[2]{Aurelio F. Bariviera \thanks{aurelio.fernandez@urv.cat}}
\author[1]{Alejandro Perez-Laborda}
\affil[1]{\scriptsize  Universitat Rovira i Virgili, Department d'Economia, ECO-SOS, Reus, Spain}
\affil[2]{\scriptsize  Universitat Rovira i Virgili, Department of Business, Reus, Spain}

\begin{document}
\maketitle

\begin{abstract}
This paper studies the dynamic market linkages among cryptocurrencies during August 2015 - July 2020 and finds a substantial increase in market linkages for both returns and volatilities.  We use different methodologies to check the different aspects of market linkages. Financial and regulatory implications are discussed.

\end{abstract}


\section{Introduction \label{sec:intro}}
Bitcoin, the first-ever created cryptocurrency, stems from a proposal to bypass the established financial system to make peer-to-peer payments. Such initiative, born in the aftermath of the financial crisis of 2007-2008, was funneled using an anonymously posted white paper \citep{Nakamoto}. The initial popularity of Bitcoin came from the libertarian point of view, advocating for a competition of this peer-to-peer system with fiat money to overcome national boundaries. Despite its impracticability as a means of payment\footnote{According to \cite{coinmap}, there are only 16614 venues (cafeterias, groceries, ATMs, etc.) in the world that accept Bitcoin as means of payment.}, market participants soon acknowledged it as a new investment instrument. Popularity has been gaining momentum and stimulating the creation of new alternative cryptocurrencies (altcoins). Litecoin was created in 2011, Ripple in 2013, Dash, NEO, and Monero in 2014, to name a few. As of July 2020, there are more than 5000 cryptocurrencies, with a total market capitalization of \$B 344. Even though Bitcoin is still the leading player, accounting for 60\% of the market capitalization, it has been losing market participation, especially since June 2017 \citep{coinmarketcap}.

Cryptocurrencies have been consolidated as an alternative investment to traditional assets. Considering the increase in the number of coins, some cryptocurrency development companies begin to design cryptocurrency indices to monitor the evolution of the market. Also, some investors have started considering investing in portfolios predominantly constituted by different cryptocurrencies. In light of these recent market developments, this paper aims at examining the dynamic market linkages in cryptocurrencies for both returns and volatility. We believe our findings have important implications for active diversification strategies in portfolios consisting of several cryptocurrencies, and for prudential regulation regarding the stability of the market. 
The paper tackles these questions by studying the evolution of return and volatility linkages across cryptocurrencies since 2015. The contribution to the existing literature is multiple. First, we employ an extended sample that includes recent data up to July 2020, thus covering recent events that might have largely affected these markets, such as the COVID-19 pandemic. Second, we adopt different econometric methodologies to cross-check the relevance of our findings. Third, we assess the linkages across frequency-ranges, which allow us to distinguish whether the transmitted shocks across cryptocurrencies have short or long-run effects. This distinction is crucial to interpret connectedness in terms of systemic risk because market participants have different preferences over trading horizons.\footnote{Consider, for example, that cryptocurrencies were tight at high frequencies only. In such a situation, transmitted shocks are not persistent, having short-term effects only. As a result, interdependencies would not be much of an issue for an agent looking for long-run profits but would matter for a short-term trader.}

The rest of the paper is organized as follows: section \ref{sec:literature} conducts a brief literature review; section \ref{sec:methods} outlines the methodology used in the paper; section \ref{sec:data} describes data and discusses the main results; finally, section \ref{sec:conclusions} summarize the main conclusions. 

\section{Literature review \label{sec:literature}}
The early financial literature on cryptocurrencies mainly focuses on the assessment of the informational efficiency of Bitcoin. \cite{Urquhart2016} employs autocorrelation, runs, and variance ratio tests over 2010 -2016 and finds that although Bitcoin showed signs of inefficiency in the initial period of 2010-2013, it evolved towards a more efficient market later on. Shortly afterward, \cite{Bariviera2017} shows that although returns have become more efficient over time, volatility still exhibits substantial persistence. \cite{TIWARI2018106} later confirms these results. 

A different stream of the literature studies the relationships across cryptocurrencies, or between cryptocurrencies and traditional assets. \cite{CORBET201828}, for example,  provides evidence of a relative detachment of Bitcoin, Ripple, and Litecoin, from stocks, government bonds, and gold indices, thus offering some diversification benefits for investors in the short term. In a similar vein, \cite{Aslanidis2019} find a positive but time-varying conditional correlation among cryptocurrencies (Bitcoin, Ripple, Dash, Monero), and confirm their negligible relationship with traditional assets. Additionally, \cite{VIDALTOMAS-Herding2019} finds evidence of herding behavior during down markets, and that the smallest coins follow the path of the larger ones (not only that of Bitcoin). In a more recent paper \cite{BOURI2020101497} reports that the average return equicorrelation between cryptocurrencies is upward trending, which suggests that market linkages are increasing over time.  \cite{Kurka2019} argues that although Bitcoin seems isolated from other financial assets over the entire period, market linkages arise when sub-periods are carefully examined.

Perhaps, the most closely related works are \cite{YI201898} and \cite{JI2019257}. These two studies are based on the connectedness methodology of \cite{Diebold2009,Diebold2012,Diebold2014} and quantify the interdependencies across cryptocurrencies using data up to 2018. In particular, \cite{YI201898} analyzes return and volatility connectedness between six leading cryptocurrencies, stressing the importance of Bitcoin and Litecoin as sources of uncertainty. \cite{JI2019257} focuses only on volatility linkages using a large set of coins.The authors document a period of increasing interdependencies across volatilities from mid- 2016, emphasizing the role of cryptocurrencies other than Bitcoin in emitting uncertainty. 

We complement these two interesting studies in two ways. First, we study both return and volatility linkages adopting a more general set of methodologies and a broader time coverage (up to July 2020), thus capturing the most recent events. In addition, we  quantify the market linkages across frequency ranges, determining the specific frequencies at which cryptocurrencies are more tightly connected. The frequency-domain analysis allows us to document new stylized facts about the cyclical properties of the transmission mechanism, which are essential to make an overall assessment of systemic risk.

For the sake of brevity, we refer to two recent surveys covering most aspects of cryptocurrencies research topics \citep{Corbet2019,MeredizBariviera2019}.

\section{Methodology \label{sec:methods}}
We conduct the empirical analysis using three different methodologies to assess cryptocurrency market linkages. Our first approach, Principal Component Analysis (PCA), is a statistical method that converts a set of correlated variables into a set of uncorrelated components through an orthogonal transformation. PCA aims to reduce the dimension of the data retaining as much variance (information) as possible. See, e.g., \cite{Wei} for further details about PCA methodology. 

Our second approach to assess cryptocurrency market linkages is based on the estimation of cross-sectional dependence. Specifically, we first obtain the pair-wise cross-sectional correlations of the cryptocurrencies, $\hat{\rho}_{ij}$.  Then, we calculate the average correlations across all pairs as $\overbar{\hat{\rho}}=(2/N(N-1))\sum_{i=1}^{N-1}\sum_{j=i+1}^{N}\hat{\rho}_{ij}$, and the associated cross-sectional dependence statistic of \cite{Pesaran2015} as $CD=\left[TN(N-1)/2\right]^{1/2}\overbar{\hat{\rho}}.$ \cite{Pesaran2015} establishes that the implicit null hypothesis of the CD test is that of weak cross sectional dependence versus the alternative hypothesis of strong cross sectional dependence\footnote{For further developments on cross-sectional dependence, we refer to the Special Issue edited by \cite{BaltagiPesaran}}. 

Our third approach consists of constructing quantitative measures of market interdependence (or connectedness) based on the vector autorregression (VAR) framework of  \cite{Diebold2009,Diebold2012,Diebold2014}. This methodology has also been used by other papers in the literature \citep{YI201898, JI2019257}. Beside this, we also follow the approach of \cite{BarunikKrehlik2018}, who extend the traditional Diebold-Yilmaz framework to the frequency-domain. The advantage of the frequency-domain is twofold. First, the frequency-domain analysis allows us to distinguish whether shocks across cryptocurrencies have long- or short-term effects. Second, one can recover standard, (time-domain) indices by aggregating frequency-domain connectedness measures over all frequencies. Thus, the approach in \cite{BarunikKrehlik2018} allows for a simultaneous assessment over time and across frequencies. \cite{CORBET201828} for example, also rely on this approach when assessing connectedness between cryptocurrencies and traditional assets. We briefly discuss the major features of the frequency-domain measures in the supplementary material. However, for the standard time-domain indices, we refer to \cite{Diebold2012, Diebold2014}.

\section{Data and results \label{sec:data}}
The empirical analysis employs daily data for seventeen major cryptocurrencies obtained from \url{https://coinmarketcap.com}. Since these cryptocurrencies were not launched at the same time, to capture more information we expand our sample adding more coins as we move through in time. Sample 1 includes seven important cryptocurrencies traded since (at least) August 2015. Sample 2 adds three more cryptocurrencies with data starting in October 2016. Finally, Sample 3 adds seven more coins, and data coverage starts in October 2017. Returns are computed as the log-price differences. We follow \cite{Diebold2012}, and estimate daily range-based return volatilities from open, close, high, and low daily prices, as in \cite{Garman}. Given that realized volatilities are right-skewed but approximately Gaussian after taking logs \citep{Anders}, we consider logarithmically transformed volatilities as time series for our estimations, as in \cite{Diebold2016} or \cite{Dem}. 

\begin{table}[!htbp]
  \centering
  \caption{Factor analysis computed with different number of coins in each dataset and in non-overlapping 1-year windows.}
    \resizebox{.99\textwidth}{!}{
  \begin{tabular}{clrrrrrrrrrrrrrrrrrrrrrr}
\cmidrule{3-12}    \multicolumn{2}{l}{\textbf{Sample 1 (7 coins)}} & \multicolumn{5}{c|}{Returns}          & \multicolumn{5}{c}{Volatilities} \\
          &       & 08/08/2015 & 08/08/2016 & 08/08/2017 & 08/08/2018 & \multicolumn{1}{r|}{08/08/2019} & 08/08/2015 & 08/08/2016 & 08/08/2017 & 08/08/2018 & 08/08/2019 \\
          &       & 07/08/2016 & 07/08/2017 & 07/08/2018 & 07/08/2019 & \multicolumn{1}{r|}{17/07/2020} & 07/08/2016 & 07/08/2017 & 07/08/2018 & 07/08/2019 & 17/07/2020 \\
\cmidrule{3-12}    \multicolumn{1}{c}{\% Variance explained by first PC} & \ & 36\%  & 37\%  & 62\%  & 80\%  & 83\%  & 41\%  & 54\%  & 70\%  & 73\%  & 71\% \\
    \multirow{7}[1]{*}{Squared component loading} & BTC   & 76\%  & 49\%  & 60\%  & 79\%  & 86\%  & 61\%  & 63\%  & 76\%  & 76\%  & 67\% \\
          & DASH  & 21\%  & 39\%  & 59\%  & 78\%  & 65\%  & 36\%  & 49\%  & 69\%  & 74\%  & 69\% \\
          & ETH   & 7\%   & 35\%  & 80\%  & 88\%  & 93\%  & 27\%  & 66\%  & 76\%  & 85\%  & 83\% \\
          & LTC   & 68\%  & 36\%  & 68\%  & 79\%  & 91\%  & 43\%  & 67\%  & 75\%  & 74\%  & 83\% \\
          & XLM   & 27\%  & 36\%  & 49\%  & 79\%  & 74\%  & 40\%  & 55\%  & 58\%  & 64\%  & 48\% \\
          & XMR   & 31\%  & 41\%  & 70\%  & 83\%  & 85\%  & 39\%  & 13\%  & 71\%  & 76\%  & 72\% \\
          & XRP   & 19\%  & 19\%  & 50\%  & 70\%  & 87\%  & 37\%  & 61\%  & 67\%  & 63\%  & 74\% \\
    \midrule
          &       &       &       &       &       &       &       &       &       &       &  \\
\cmidrule{4-7}\cmidrule{9-12}    \multicolumn{2}{l}{\textbf{Sample 2 (10 coins)}} &       & \multicolumn{4}{c}{Returns}   &       & \multicolumn{4}{c}{Volatilities} \\
          &       &       & 30/10/2016 & 30/10/2017 & 30/10/2018 & 30/10/2019 &       & 30/10/2016 & 30/10/2017 & 30/10/2018 & 30/10/2019 \\
          &       &       & 29/10/2017 & 29/10/2018 & 29/10/2019 & 17/07/2020 &       & 29/10/2017 & 29/10/2018 & 29/10/2019 & 17/07/2020 \\
\cmidrule{4-7}\cmidrule{9-12}    \multicolumn{1}{c}{\%Variance explained by first PC} &  \ &     & 37\%  & 65\%  & 79\%  & 82\%  &       & 48\%  & 75\%  & 72\%  & 72\% \\
    \multirow{10}[0]{*}{Squared component loading} & BTC   &       & 52\%  & 56\%  & 79\%  & 87\%  &       & 55\%  & 79\%  & 76\%  & 64\% \\
          & DASH  &       & 44\%  & 62\%  & 80\%  & 66\%  &       & 50\%  & 76\%  & 72\%  & 75\% \\
          & ETC   &       & 52\%  & 64\%  & 72\%  & 79\%  &       & 54\%  & 73\%  & 74\%  & 73\% \\
          & ETH   &       & 50\%  & 80\%  & 89\%  & 92\%  &       & 67\%  & 81\%  & 83\%  & 81\% \\
          & LTC   &       & 46\%  & 67\%  & 78\%  & 91\%  &       & 57\%  & 80\%  & 72\%  & 85\% \\
          & NEO   &       & 26\%  & 66\%  & 80\%  & 84\%  &       & 24\%  & 78\%  & 78\%  & 65\% \\
          & XLM   &       & 27\%  & 58\%  & 76\%  & 77\%  &       & 49\%  & 70\%  & 55\%  & 55\% \\
          & XMR   &       & 48\%  & 74\%  & 82\%  & 86\%  &       & 45\%  & 78\%  & 68\%  & 75\% \\
          & XRP   &       & 13\%  & 52\%  & 78\%  & 87\%  &       & 54\%  & 61\%  & 68\%  & 76\% \\
          & ZEC   &       & 16\%  & 74\%  & 79\%  & 72\%  &       & 25\%  & 69\%  & 71\%  & 70\% \\
          &       &       &       &       &       &       &       &       &       &       &  \\
\cmidrule{5-7}\cmidrule{10-12}    \multicolumn{2}{l}{\textbf{Sample 3 (17 coins)}} &       &       & \multicolumn{3}{c}{Returns} &       &       & \multicolumn{3}{c}{Volatilities} \\
          &       &       &       & 03/10/2017 & 03/10/2018 & 03/10/2019 &       &       & 03/10/2017 & 03/10/2018 & 03/10/2019 \\
          &       &       &       & 02/10/2018 & 02/10/2019 & 17/07/2020 &       &       & 02/10/2018 & 02/10/2019 & 17/07/2020 \\
\cmidrule{5-7}\cmidrule{10-12}    \multicolumn{1}{c}{\% Variance explained by first PC} & \ &       &       & 54\%  & 74\%  & 81\%  &       &       & 54\%  & 74\%  & 81\% \\
    \multirow{17}[0]{*}{Squared component loading} & ADA   &       &       & 47\%  & 85\%  & 84\%  &       &       & 47\%  & 85\%  & 84\% \\
          & BCH   &       &       & 47\%  & 69\%  & 88\%  &       &       & 47\%  & 69\%  & 88\% \\
          & BNB   &       &       & 38\%  & 54\%  & 89\%  &       &       & 38\%  & 54\%  & 89\% \\
          & BTC   &       &       & 54\%  & 76\%  & 87\%  &       &       & 54\%  & 76\%  & 87\% \\
          & DASH  &       &       & 59\%  & 78\%  & 63\%  &       &       & 59\%  & 78\%  & 63\% \\
          & EOS   &       &       & 54\%  & 80\%  & 90\%  &       &       & 54\%  & 80\%  & 90\% \\
          & ETC   &       &       & 63\%  & 72\%  & 78\%  &       &       & 63\%  & 72\%  & 78\% \\
          & ETH   &       &       & 78\%  & 89\%  & 93\%  &       &       & 78\%  & 89\%  & 93\% \\
          & LTC   &       &       & 62\%  & 79\%  & 92\%  &       &       & 62\%  & 79\%  & 92\% \\
          & MIOTA &       &       & 58\%  & 71\%  & 75\%  &       &       & 58\%  & 71\%  & 75\% \\
          & NEO   &       &       & 63\%  & 84\%  & 78\%  &       &       & 63\%  & 84\%  & 78\% \\
          & TRX   &       &       & 34\%  & 68\%  & 87\%  &       &       & 34\%  & 68\%  & 87\% \\
          & XLM   &       &       & 44\%  & 76\%  & 76\%  &       &       & 44\%  & 76\%  & 76\% \\
          & XMR   &       &       & 71\%  & 80\%  & 84\%  &       &       & 71\%  & 80\%  & 84\% \\
          & XRP   &       &       & 49\%  & 77\%  & 86\%  &       &       & 49\%  & 77\%  & 86\% \\
          & XTZ   &       &       & 23\%  & 45\%  & 64\%  &       &       & 23\%  & 45\%  & 64\% \\
          & ZEC   &       &       & 70\%  & 77\%  & 69\%  &       &       & 70\%  & 77\%  & 69\% \\
    \end{tabular}%
    }
  \label{tab:PCA}%
\end{table}%

We first conduct PCA for returns and volatilities, separately. Following the literature, we standardize the data before applying PCA to prevent undue influence of a variable. We provide comparative analysis, dividing the data into non-overlapping one-year samples. The percentage of the variance explained by the first principal component is reported in Table \ref{tab:PCA}. The table also provides the squared component loadings, which are just the squared correlations between the first principal component and each of the variables. Analogous to Pearson's R-squared, the squared component loading measures the percentage of the variance in that variable explained by the principal component.

Results in Table \ref{tab:PCA} illustrate the increasing commonality in return variances in the cryptocurrency market. The first principal component (PC) explains 36\% of the overall variance in the first year of Sample 1. It is important to point out that in this period, the first PC represents around 70\% of Bitcoin or Litecoin variance. However, for other coins such as Ethereum, the first PC would capture only 7\% of variance. This picture changes dramatically over time. In the last yearly sample (August 2019 - July 2020) the first principal component explains more than 80\% of the overall data variance and at least 65\% of the variance of any individual coin. Thus, the PCA of returns reflects a clear path toward closer cryptocurrency linkages, where shocks are rapidly transmitted across the market. Overall, we conclude that one PC is now sufficient to represent the dynamics of the entire cryptocurrency market. 

Table \ref{tab:PCA} also reports PCA for the volatility series. The results are similar to those obtained for returns. Volatility linkages increase substantially over time, albeit towards the end of the sample period reaching slightly lower levels of interdependency than returns.

\begin{figure}[!htbp]
\centering
\includegraphics[width = 0.6\textwidth]{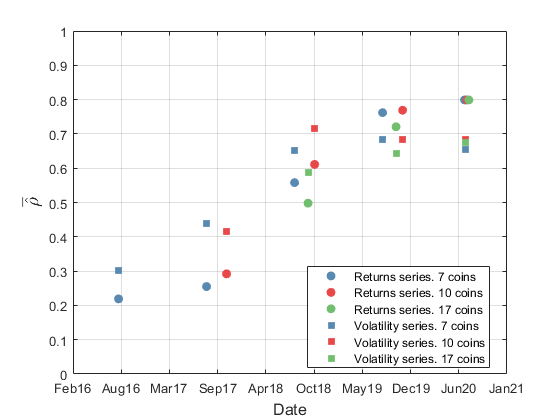} 
\caption{Pesaran cross-sectional dependence, average correlations $\overbar{\hat{\rho}}$, computed using different number of coins in each sample and in non-overlapping 1-year windows. Coins and windows dates are the same as in Table \ref{tab:PCA}}
\label{fig:Pesaran}
\end{figure}

Next, we conduct the \cite{Pesaran2015} test for cross-sectional dependence. Our results strongly reject the null hypothesis of weak in favor of strong cross-sectional dependence in all cases\footnote{Detailed results of CD test are not reported in the paper, but they are available upon request.}. The average correlation across all pairs (for returns and volatilities, separately) is plotted in Figure \ref{fig:Pesaran}. The increasing cross-sectional correlations over time are consistent with previous findings obtained using PCA. Again, towards the end of the sample the increase in correlation is slightly stronger in returns than in volatilities. Thus, our results support previous findings of steadily increasing cross-cryptocurrency market linkages.


Finally, to quantify the strength of these market linkages we carry out a more extensive linkage analysis using the methodologies developed in \cite{Diebold2009,Diebold2012,Diebold2014} and \cite{BarunikKrehlik2018}. For each dataset, we analyze the dynamic evolution of connectedness estimating VARs for returns and for volatilities in a rolling window fashion.\footnote{The results in the paper are based on a vector autoregression of order four and a rolling window length of 365 days (one year)} As in \cite{BarunikKrehlik2018}, the usual (time-domain) connectedness indices are computed by aggregating frequency connectedness over all ranges, but the results are identical to those obtained from finite-horizon formulas with the standard ten period ahead horizon used, for example, in \cite{Diebold2012}.

\begin{figure}[!htbp]
\centering
\subfloat[Returns]{\includegraphics[width = 0.6\textwidth]{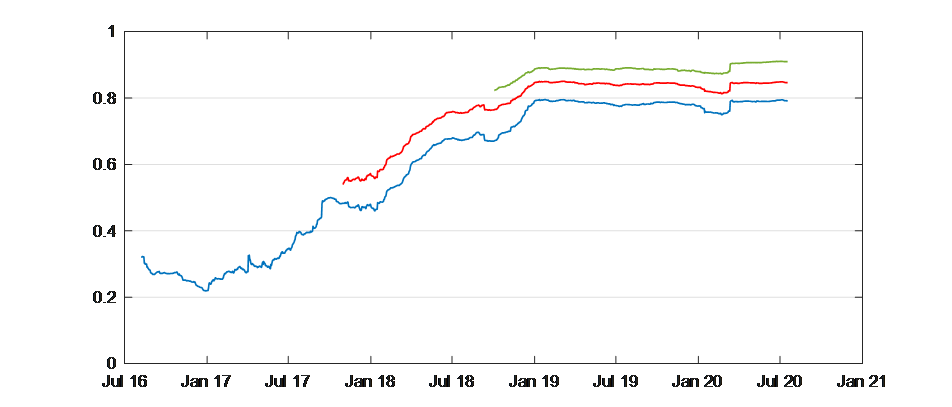} \label{fig:TotalConnecteness-A}}\\ 
\subfloat[Volatilities]{\includegraphics[width=0.6\textwidth]{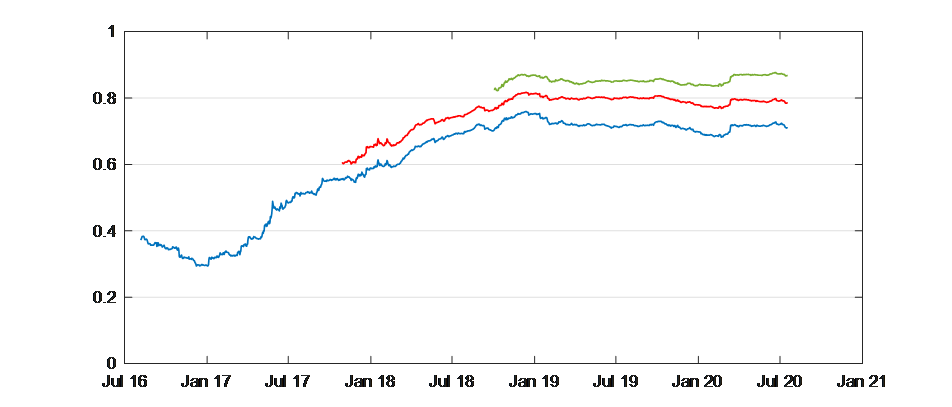}\label{fig:TotalConnecteness-B}}\\
\subfloat{\includegraphics[width = 0.6\textwidth]{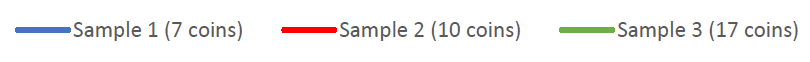}}\\
\caption{Total connectedness in returns and volatilties, using rolling windows.}
\label{fig:TotalConnecteness}
\end{figure}

Figure \ref{fig:TotalConnecteness} depicts the dynamic evolution of the total connectedness indices for returns (upper panel) and for volatilities (bottom panel) using the three samples of coins. Total connectedness (also called spillover index) measures, on average, the percentage of the variance explained by shocks transmitted across coins. As Figure \ref{fig:TotalConnecteness} shows, total connectedness exhibits a substantial upward trend for both returns and volatilities. Specifically, the index for returns rises from nearly 25\% to 80\% over 2016-2020 while for volatilities from roughly 30\% to 76\% during the same period. Thus, more than three quarters of the return and volatility variances are currently explained by shocks transmitted across cryptocurrencies. Notice that, despite the different starting dates, the magnitude and pattern of total connectedness are very similar across the three samples, that is, our findings are robust and therefore independent of the number of coins considered.

Since 2019 total connectedness stabilizes around 0.8, leaving little room for further increases. Importantly, we observe a sudden upward jump around March 2020 in all plots, which is contemporaneous to stock market crashes worldwide arising from the COVID-19 pandemic. This result indicates that such a global shock has helped to knit even tighter relationships among cryptocurrencies. This finding broadens the results by \cite{Goodell2020} regarding the influence of COVID-19 on Bitcoin prices.

\begin{table}[!htbp]
  \centering
  \caption{ Return connectedness table for Sample 1 (7 coins)}
\resizebox{.7\textwidth}{!}{
  \begin{tabular}{lrrrrrrrc}
   \multicolumn{9}{c}{Period 08/08/ 2015 - 07/08/2016. Total connectedness: 0.32} \\
\toprule           & \multicolumn{1}{l}{BTC} & \multicolumn{1}{l}{DASH} & \multicolumn{1}{l}{ETH} & \multicolumn{1}{l}{LTC} & \multicolumn{1}{l}{XLM} & \multicolumn{1}{l}{XMR} & \multicolumn{1}{l}{XRP} & \multicolumn{1}{l}{Total from} \\
\cmidrule{2-9}    BTC   & 0.47  & 0.05  & 0.01  & 0.30  & 0.05  & 0.09  & 0.04  & \multicolumn{1}{r}{0.53} \\
    DASH  & 0.08  & 0.79  & 0.01  & 0.06  & 0.02  & 0.02  & 0.02  & \multicolumn{1}{r}{0.21} \\
    ETH   & 0.02  & 0.02  & 0.88  & 0.01  & 0.02  & 0.03  & 0.01  & \multicolumn{1}{r}{0.12} \\
    LTC   & 0.32  & 0.04  & 0.01  & 0.52  & 0.05  & 0.04  & 0.04  & \multicolumn{1}{r}{0.48} \\
    XLM   & 0.09  & 0.02  & 0.02  & 0.08  & 0.67  & 0.03  & 0.09  & \multicolumn{1}{r}{0.33} \\
    XMR   & 0.13  & 0.03  & 0.03  & 0.07  & 0.03  & 0.70  & 0.02  & \multicolumn{1}{r}{0.30} \\
    XRP   & 0.05  & 0.02  & 0.02  & 0.05  & 0.11  & 0.02  & 0.73  & \multicolumn{1}{r}{0.27} \\
 \cmidrule{1-8}     Total to & 0.69  & 0.17  & 0.10  & 0.57  & 0.27  & 0.23  & 0.21  &  \\
\bottomrule          &       &       &       &       &       &       &       &  \\
   \multicolumn{9}{c}{Period 19/07/2019 - 18/07/2020. Total connectedness: 0.79} \\
\toprule           & \multicolumn{1}{l}{BTC} & \multicolumn{1}{l}{DASH} & \multicolumn{1}{l}{ETH} & \multicolumn{1}{l}{LTC} & \multicolumn{1}{l}{XLM} & \multicolumn{1}{l}{XMR} & \multicolumn{1}{l}{XRP} & \multicolumn{1}{l}{Total from} \\
\cmidrule{2-9}    BTC   & 0.20  & 0.09  & 0.17  & 0.15  & 0.10  & 0.15  & 0.13  & \multicolumn{1}{r}{0.80} \\
    DASH  & 0.12  & 0.26  & 0.14  & 0.14  & 0.10  & 0.13  & 0.12  & \multicolumn{1}{r}{0.74} \\
    ETH   & 0.15  & 0.10  & 0.19  & 0.16  & 0.11  & 0.14  & 0.15  & \multicolumn{1}{r}{0.81} \\
    LTC   & 0.15  & 0.10  & 0.16  & 0.19  & 0.12  & 0.14  & 0.15  & \multicolumn{1}{r}{0.81} \\
    XLM   & 0.12  & 0.08  & 0.14  & 0.14  & 0.22  & 0.13  & 0.16  & \multicolumn{1}{r}{0.78} \\
    XMR   & 0.15  & 0.10  & 0.15  & 0.14  & 0.11  & 0.21  & 0.13  & \multicolumn{1}{r}{0.79} \\
    XRP   & 0.13  & 0.09  & 0.16  & 0.15  & 0.14  & 0.13  & 0.20  & \multicolumn{1}{r}{0.80} \\
\cmidrule{1-8}      Total to & 0.82  & 0.57  & 0.92  & 0.89  & 0.68  & 0.82  & 0.84  &  \\
\bottomrule    
\end{tabular}%
}
  \label{tab:pairwise}%
\end{table}%

The directional connectedness indices offer further insight into the evolution of the cryptocurrency market linkages. In Table \ref{tab:pairwise} we report the complete connectedness tables for returns (also known as spillover table) estimated using the first and last years of the seven coins sample (Sample 1). Results in Table \ref{tab:pairwise} provide strong evidence of an across the board increase in return connectedness already depicted in Figure \ref{fig:TotalConnecteness}. Moreover, a comparison of the pairwise and the total FROM and TO connectedness indices shows that the relative importance of the different cryptocurrencies have also changed dramatically over time. During the first year, spillovers are mostly driven by Bitcoin and Litecoin, while other coins seem less interconnected. This result is consistent with the findings in \cite{YI201898} using data up to February 2018. However, the situation changes remarkably in recent years, since all coins are now about equally important in transmitting/receiving return spillovers (shocks across markets represent between 74\% to 80\% of the return variances of coins).

The directional connectedness analysis of the volatility linkages is similar to that of returns. See supplementary material for detailed results.\footnote{The supplement also provides connectedness tables obtained with the seventeen coins sample (Sample 3)}

Overall, our results show that cryptocurrencies have become increasingly interconnected in both return and volatility over the recent period, emerging as a compactly integrated market. These results complement the previous findings on volatility in \cite{JI2019257} and highlight that the cryptocurrency market is becoming significantly more vulnerable to within shock transmissions.

\begin{figure}[!htbp]
\centering
\includegraphics[width = 0.4\textwidth]{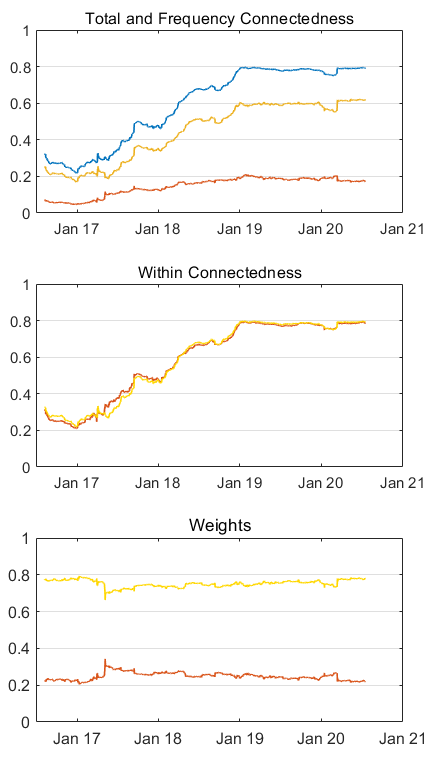}
\includegraphics[width = 0.385\textwidth]{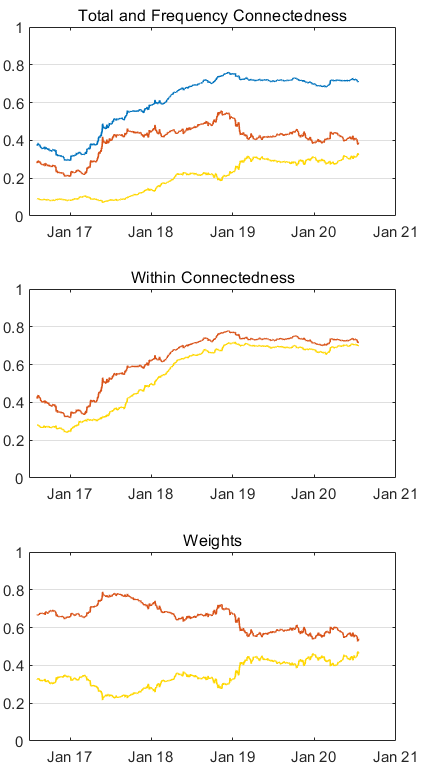}\\
\includegraphics[width = 0.6\textwidth]{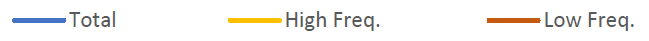}
\caption{Connectedness across frequency ranges for Sample 1 of seven coins (Returns: left panels;  Volatilities: right panels). High frequencies: 1-7 day period . Low frequencies: period longer than 7 days.}
\label{fig:h-lConnecteness-1}
\end{figure}

We further adopt the \cite{BarunikKrehlik2018} approach to evaluate connectedness in the frequency domain (high vs. low frequencies). The frequency domain approach would help us disentangle the specific frequencies that have most contributed to the observed rise in connectedness. The high-frequency range includes frequencies with periods from one to seven days (one week), while the low-frequency range frequencies with periods longer than one week. To the best of our knowledge, the frequency domain method is not performed in the other studies in the literature focusing on the existing relationships across cryptocurrencies, such as \cite{YI201898} and \cite{JI2019257}.   

Figure \ref{fig:h-lConnecteness-1} plots connectedness measures by frequency ranges obtained using the seven-coin sample (Sample 1)-- corresponding figures for the other samples with more coins can be found in the Supplement. The upper panel of \ref{fig:h-lConnecteness-1} depicts the decomposition of the total connectedness index in \ref{fig:TotalConnecteness} into two frequency connectedness components: the high frequency and low frequency components. Notice that connectedness at the two frequency ranges add up the total connectedness index. The middle panel plots the connectedness created within the specific ranges. Finally, the bottom panel plots the weights used to transform within connectedness into the frequency connectedness components, which measures the relative importance of high and low fluctuations on total variance.

Regarding returns (left panels), most of the connectedness between returns is created at high frequencies. The high-frequency component accounting for most of the observed increase in the total connectedness index. This implies that shocks across cryptocurrencies have mostly temporary effects on returns, dissipating fast in the short-run. This result, however, does not imply that returns are not connected at low frequencies. In fact, as the figures in the second and third panel show, returns are about equally connected at high and low frequencies, but fluctuations at high frequencies turn out to be more important for returns. Besides, it is suggestive that in March 2020, the relative importance of high-frequency fluctuations experiences a sudden increase, which indicates that the COVID-19 pandemic has increased the importance of immediacy. 
The relative small importance of low-frequency connection supports previous findings regarding long-range memory (e.g., \cite{Urquhart2016, Bariviera2017, TIWARI2018106}).

The results for volatility differ substantially from those of returns. As the right panels of Figure \ref{fig:h-lConnecteness-1} show, most of the volatility connectedness is created at low frequencies. This indicates that volatility shocks across cryptocurrencies have persistent effects. However, we also observe an increasing contribution of the high-frequency component over time, which indicates that that information is currently transmitted faster across cryptocurrencies than it was before. As the middle and lower panels of the figure show, most of this increase can be explained by a strong decline in the relative importance of low-frequency fluctuations on the volatility variances. This decline may signal that nowadays agents are  better able to offset any long-run effects of shocks by switching to other assets. In any case, these are good news for long-run investors, as their exposure to systemic risk over the long-term has significantly decreased.
\section{Conclusions \label{sec:conclusions}}

This paper broadens previous studies on cryptocurrency market linkages. We tackle this issue by an ensemble of methodologies to examine return and volatility linkages across the major coins over the last five years. To account for the fact that new coins are being introduced in the market, we conduct our analysis using extended samples with an increasing number of coins. 
Irrespective of the methodology adopted, we document that the cryptocurrency market has experienced a strong overall increase in market linkages (return and volatility). As of July 2020, only few coin-specific shocks are not transmitted to the rest of the coins (less than 20\%).
The insights provided by the frequency-domain approach have provided new stylized facts on the shock transmission mechanism across cryptocurrencies. The paper uncovers that the transmitted shocks have mostly short-term effects on returns. This result is in line with the view that the cryptocurrency market makes significant steps towards becoming efficient. Although a significant part of volatility connectedness is still created at low-frequencies, we show that volatility transmission at high frequencies has currently become considerably more important.

Our results have several practical implications. 
First, there are now limited diversification benefits in the cryptocurrency market, with active portfolio re-balancing becoming mostly irrelevant. Second, cryptocurrency indices hardly add any information about market evolution beyond that conveyed by any individual cryptocurrency. Third, from a regulatory perspective, if cryptocurrencies were to become legal tender at some point in time, policymakers should evaluate the potentially disruptive effects of such a highly interconnected market. Nevertheless, the observed decline of the relative importance of low-frequency transmission is favoring long-term investors in terms of smaller exposure to systemic risk.

 \section*{SUPLEMENTARY MATERIAL}  
\appendix

\section{Connectedness in the frequency domain.}

In this section, we briefly discuss the major features of the frequency domain connectedness developed in \cite{BarunikKrehlik2018}. Like in the typical, time-domain \cite{Diebold2012} framework, connectedness measures in the frequency domain are based on the generalized VAR \citep{Pes} to deal with (possibly) correlated innovations, but the authors employ the spectral decomposition of the variance instead of the forecast error variance decomposition. We refer to \cite{BarunikKrehlik2018} for further details.

Let i be the imaginary unit, and $R=\left(a,b\right):a,b\in\left(-\pi,\pi\right),a<b$ the targeted frequency range. The share of a shock to variable \textit{k} in variable \textit{j}'s fluctuations at the band \textit{R} is given by:

$$\mathrm{\Theta}_{j,k}^R=\frac{1}{2\pi}\int_{a}^{b}{P_j\left(\omega\right)f_{j,k}\left(\omega\right)d\omega},$$

The terms $f_{j,k}\left(\omega\right)$ and $P_j\left(\omega\right)$ are the generalized causation (cross-)spectrum of the VAR and the power of variable \textit{j} at frequency $\omega$, and the definite integral can be approximated by summations for Fourier frequencies $\omega_j=2\pi j/T$, $j=1,\dots T/2$ belonging to the frequency range \textit{R}. Like in the time-domain, these shares are normalized as  ${\widetilde{\mathrm{\Theta}}}_{j,k}^d=\mathrm{\Theta}_{j,k}^R/{\sum_{k}\mathrm{\Theta}_{j,k}^\infty}$, where $\mathrm{\Theta}_{j,k}^\infty$ denotes the contribution over all frequencies, and arranged after in a matrix of normalized contributions. The \textit{within connectedness} at the frequency range \textit{R} is defined from these normalized contributions as:

$$WC^R=1-\frac{Tr\left\{{\widetilde{\mathrm{\Theta}}}^R\right\}}{\sum{\widetilde{\mathrm{\Theta}}}^R}.$$

Within connectedness quantifies the contribution of shock transmission on the fluctuations at the specific range of frequencies \textit{R}, on average in the system. However, this index does not account for the relative importance of these fluctuations on the total system variance. As a result, overall connectedness can be low even if within connectedness is high at a particular range. To account for this, within connectedness is weighted by the relative importance of the band to define \textit{frequency connectedness} at the range:
	
	 $$FC^R=WC^R\frac{\sum{\widetilde{\mathrm{\Theta}}}^R}{\sum{\widetilde{\mathrm{\Theta}}}^\infty}$$
	
The authors show that frequency connectedness decomposes total (time-domain) connectedness in components at different ranges. Specifically, let ${C^h}$ be the DY total connectedness index at a forecast horizon $h$ (i.e., the spillover index). Consider a set of ranges $R_s$ that form a partition of the space $(-\pi,\pi)$. Then it can be shown that:

$$\lim_{h\to\infty}{C^h}=\sum_{R_s}{FC^{R_s}}$$

Notice that equality holds on the limit. However, VAR variance decomposition typically converge fast, and the previous equation delivers very good approximations for finite horizons as well, provided those are not too short. 

\subsection{Connectedness across frequency ranges for Samples 2 and 3}

Below we provide return and volatility connectedness measures for the high and for the low frequency ranges using Sample 2 (10 coins) and Sample 3 (17 coins). Corresponding figures for Sample 1 are reported in the main text.

\begin{figure}[H]
\centering
\includegraphics[width = 0.4\textwidth]{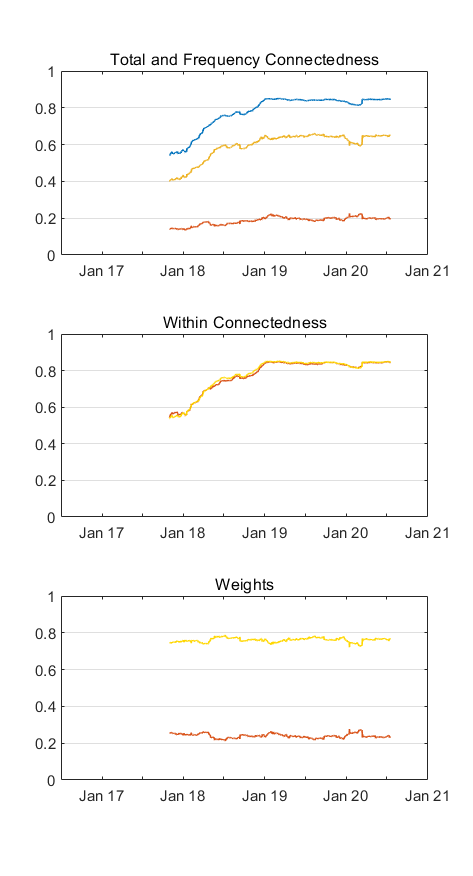}
\includegraphics[width = 0.4\textwidth]{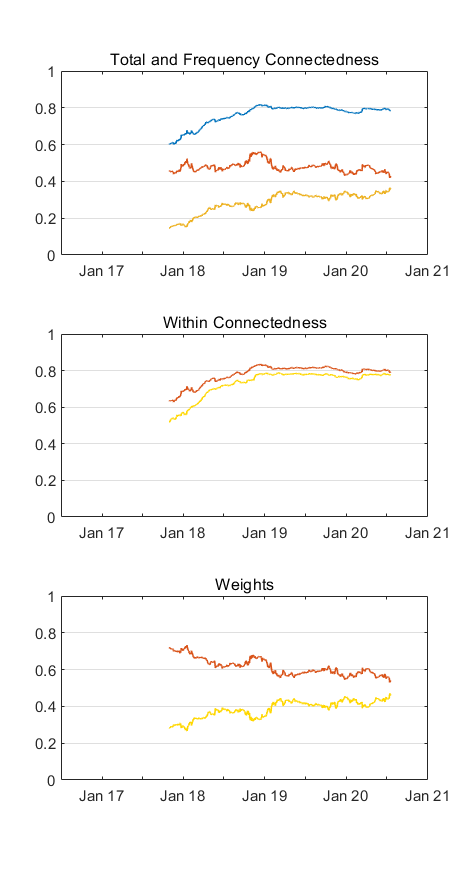}\\
\includegraphics[width = 0.40\textwidth]{legend-freq.PNG}
\caption{Connectedness across frequency ranges for Sample 2 with 10 coins (Returns: left panels;  Volatilities: right panels). High frequencies: 1-7 day period. Low frequencies: period longer than 7 days.}
\label{fig:h-lConnecteness-2}
\end{figure}

\begin{figure}[H]
\centering
\includegraphics[width = 0.4\textwidth]{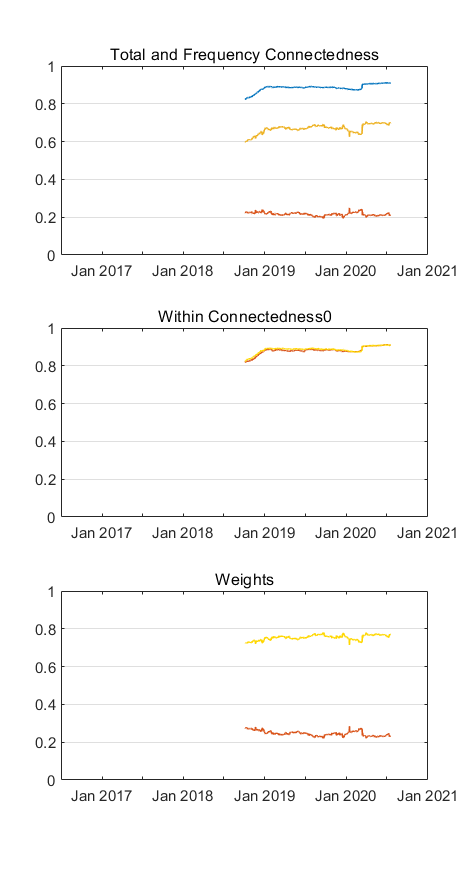}
\includegraphics[width = 0.4\textwidth]{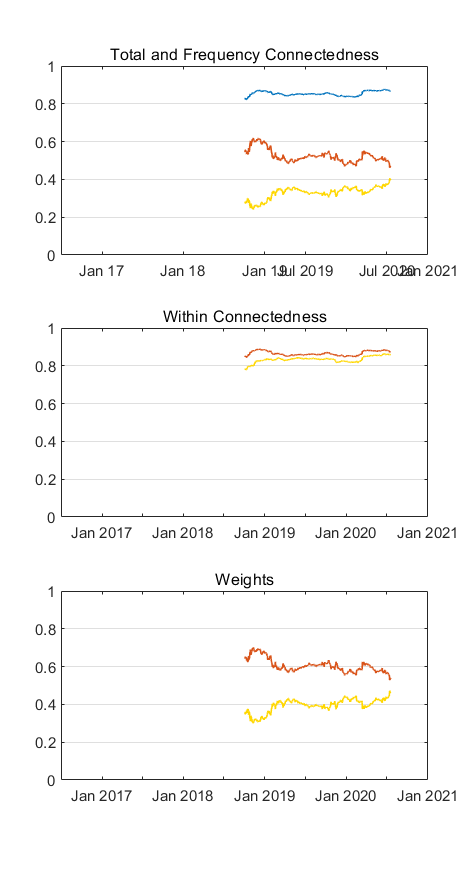}\\
\includegraphics[width = 0.4\textwidth]{legend-freq.PNG}
\caption{Connectedness across frequency ranges for Sample 3 with 17 coins (Returns: left panels;  Volatilities: right panels). High frequencies: 1-7 day period. Low frequencies: period longer than 7 days.}
\label{fig:h-lConnecteness-3}
\end{figure}

\section{ Connectedness tables at selected subsamples}

Below we provide tables with results for: (i) Volatility connectedness for Sample 1 (Table \ref{tab:pairwise7coinsvolatility}) computed using the first year and the last years of the sample period. The corresponding tables for returns are provided in the main text; (ii) Return (Table \ref{tab:pairwise17coinsreturn}) and volatility (Table \ref{tab:pairwise17coinsvolatility}) connectedness for Sample 3 computed using the last year of the sample period.

\begin{table}[H]
  \centering
  \caption{Pairwise volatility connectedness for sample 1}
      \resizebox{.6\textwidth}{!}{
    \begin{tabular}{lrrrrrrrr}
    \multicolumn{9}{c}{Period 08/08/ 2015 - 07/08/2016. Total connectedness: 0.32} \\
          & \multicolumn{1}{l}{BTC} & \multicolumn{1}{l}{DASH} & \multicolumn{1}{l}{ETH} & \multicolumn{1}{l}{LTC} & \multicolumn{1}{l}{XLM} & \multicolumn{1}{l}{XMR} & \multicolumn{1}{l}{XRP} & \multicolumn{1}{l}{Total from} \\
\cmidrule{2-9}    BTC   & 0.53  & 0.05  & 0.04  & 0.26  & 0.02  & 0.04  & 0.05  & 0.47 \\
    DASH  & 0.10  & 0.72  & 0.05  & 0.05  & 0.01  & 0.05  & 0.02  & 0.28 \\
    ETH   & 0.05  & 0.09  & 0.73  & 0.03  & 0.04  & 0.05  & 0.01  & 0.27 \\
    LTC   & 0.31  & 0.03  & 0.04  & 0.55  & 0.01  & 0.03  & 0.03  & 0.45 \\
    XLM   & 0.09  & 0.06  & 0.06  & 0.05  & 0.64  & 0.06  & 0.05  & 0.36 \\
    XMR   & 0.09  & 0.09  & 0.11  & 0.06  & 0.04  & 0.59  & 0.01  & 0.41 \\
    XRP   & 0.10  & 0.04  & 0.05  & 0.05  & 0.09  & 0.03  & 0.64  & 0.36 \\
\cmidrule{1-8}    Total to & 0.69  & 0.17  & 0.10  & 0.57  & 0.27  & 0.23  & 0.21  &  \\
\cmidrule{1-8}          &       &       &       &       &       &       &       &  \\
    \multicolumn{9}{c}{Period 19/07/2019 - 18/07/2020. Total connectedness: 0.71} \\
          & \multicolumn{1}{l}{BTC} & \multicolumn{1}{l}{DASH} & \multicolumn{1}{l}{ETH} & \multicolumn{1}{l}{LTC} & \multicolumn{1}{l}{XLM} & \multicolumn{1}{l}{XMR} & \multicolumn{1}{l}{XRP} & \multicolumn{1}{l}{Total from} \\
\cmidrule{2-9}    BTC   & 0.28  & 0.10  & 0.15  & 0.16  & 0.06  & 0.12  & 0.12  & 0.72 \\
    DASH  & 0.10  & 0.31  & 0.13  & 0.16  & 0.08  & 0.10  & 0.10  & 0.69 \\
    ETH   & 0.14  & 0.12  & 0.25  & 0.18  & 0.08  & 0.10  & 0.13  & 0.75 \\
    LTC   & 0.12  & 0.12  & 0.16  & 0.27  & 0.08  & 0.11  & 0.13  & 0.73 \\
    XLM   & 0.08  & 0.12  & 0.13  & 0.12  & 0.35  & 0.08  & 0.13  & 0.65 \\
    XMR   & 0.12  & 0.11  & 0.12  & 0.16  & 0.07  & 0.27  & 0.13  & 0.73 \\
    XRP   & 0.10  & 0.10  & 0.14  & 0.16  & 0.10  & 0.11  & 0.29  & 0.71 \\
\cmidrule{1-8}    Total to & 0.66  & 0.67  & 0.84  & 0.95  & 0.47  & 0.63  & 0.75  &  \\
\cmidrule{1-8}    
\end{tabular}%
}
  \label{tab:pairwise7coinsvolatility}%
\end{table}%

\begin{table}[H]
  \centering
  \caption{Pairwise return connectedness for sample 3}
      \resizebox{.99\textwidth}{!}{
  \begin{tabular}{lrrrrrrrrrrrrrrrrrr}
    \multicolumn{19}{c}{Period 19/07/2019 - 18/07/2020. Total connectedness: 0.91} \\
    \cmidrule{2-19}
          & \multicolumn{1}{l}{ADA} & \multicolumn{1}{l}{BCH} & \multicolumn{1}{l}{BNB} & \multicolumn{1}{l}{BTC} & \multicolumn{1}{l}{DASH} & \multicolumn{1}{l}{EOS} & \multicolumn{1}{l}{ETC} & \multicolumn{1}{l}{ETH} & \multicolumn{1}{l}{LTC} & \multicolumn{1}{l}{MIOTA} & \multicolumn{1}{l}{NEO} & \multicolumn{1}{l}{TRX} & \multicolumn{1}{l}{XLM} & \multicolumn{1}{l}{XMR} & \multicolumn{1}{l}{XRP} & \multicolumn{1}{l}{XTZ} & \multicolumn{1}{l}{ZEC} & \multicolumn{1}{l}{Total from} \\
\cmidrule{2-19}    ADA   & 0.09  & 0.06  & 0.06  & 0.06  & 0.04  & 0.06  & 0.05  & 0.07  & 0.07  & 0.06  & 0.06  & 0.06  & 0.06  & 0.06  & 0.06  & 0.04  & 0.05  & 0.91 \\
    BCH   & 0.06  & 0.08  & 0.06  & 0.06  & 0.05  & 0.07  & 0.06  & 0.07  & 0.07  & 0.05  & 0.06  & 0.06  & 0.05  & 0.06  & 0.06  & 0.04  & 0.06  & 0.92 \\
    BNB   & 0.06  & 0.06  & 0.08  & 0.06  & 0.05  & 0.06  & 0.05  & 0.07  & 0.06  & 0.05  & 0.06  & 0.06  & 0.05  & 0.06  & 0.06  & 0.04  & 0.05  & 0.92 \\
    BTC   & 0.06  & 0.07  & 0.07  & 0.08  & 0.05  & 0.06  & 0.05  & 0.07  & 0.07  & 0.05  & 0.06  & 0.06  & 0.05  & 0.06  & 0.06  & 0.04  & 0.05  & 0.92 \\
    DASH  & 0.05  & 0.07  & 0.06  & 0.05  & 0.10  & 0.06  & 0.06  & 0.06  & 0.06  & 0.05  & 0.05  & 0.06  & 0.05  & 0.06  & 0.05  & 0.03  & 0.08  & 0.90 \\
    EOS   & 0.06  & 0.07  & 0.06  & 0.06  & 0.05  & 0.08  & 0.06  & 0.07  & 0.07  & 0.05  & 0.06  & 0.06  & 0.05  & 0.06  & 0.06  & 0.03  & 0.05  & 0.92 \\
    ETC   & 0.06  & 0.07  & 0.06  & 0.06  & 0.05  & 0.06  & 0.10  & 0.06  & 0.07  & 0.05  & 0.05  & 0.06  & 0.05  & 0.06  & 0.06  & 0.04  & 0.06  & 0.90 \\
    ETH   & 0.06  & 0.07  & 0.07  & 0.06  & 0.05  & 0.06  & 0.05  & 0.08  & 0.07  & 0.05  & 0.06  & 0.06  & 0.05  & 0.06  & 0.06  & 0.04  & 0.05  & 0.92 \\
    LTC   & 0.06  & 0.07  & 0.06  & 0.06  & 0.05  & 0.07  & 0.06  & 0.07  & 0.08  & 0.05  & 0.06  & 0.06  & 0.05  & 0.06  & 0.06  & 0.04  & 0.05  & 0.92 \\
    MIOTA & 0.06  & 0.06  & 0.07  & 0.06  & 0.05  & 0.06  & 0.06  & 0.07  & 0.06  & 0.09  & 0.06  & 0.06  & 0.05  & 0.06  & 0.06  & 0.04  & 0.05  & 0.91 \\
    NEO   & 0.06  & 0.06  & 0.06  & 0.06  & 0.04  & 0.06  & 0.05  & 0.07  & 0.07  & 0.05  & 0.09  & 0.06  & 0.05  & 0.06  & 0.06  & 0.04  & 0.05  & 0.91 \\
    TRX   & 0.06  & 0.07  & 0.06  & 0.06  & 0.04  & 0.07  & 0.05  & 0.07  & 0.07  & 0.05  & 0.06  & 0.08  & 0.05  & 0.06  & 0.07  & 0.04  & 0.05  & 0.92 \\
    XLM   & 0.06  & 0.06  & 0.06  & 0.05  & 0.04  & 0.06  & 0.05  & 0.06  & 0.06  & 0.06  & 0.06  & 0.06  & 0.10  & 0.06  & 0.07  & 0.04  & 0.05  & 0.90 \\
    XMR   & 0.05  & 0.06  & 0.07  & 0.07  & 0.05  & 0.06  & 0.05  & 0.07  & 0.06  & 0.05  & 0.05  & 0.06  & 0.05  & 0.09  & 0.06  & 0.04  & 0.05  & 0.91 \\
    XRP   & 0.06  & 0.06  & 0.06  & 0.06  & 0.04  & 0.06  & 0.05  & 0.07  & 0.07  & 0.05  & 0.06  & 0.07  & 0.06  & 0.06  & 0.09  & 0.04  & 0.05  & 0.91 \\
    XTZ   & 0.05  & 0.06  & 0.07  & 0.06  & 0.04  & 0.05  & 0.05  & 0.07  & 0.06  & 0.05  & 0.05  & 0.05  & 0.05  & 0.06  & 0.06  & 0.12  & 0.05  & 0.88 \\
    ZEC   & 0.06  & 0.06  & 0.06  & 0.05  & 0.07  & 0.06  & 0.05  & 0.07  & 0.06  & 0.05  & 0.05  & 0.06  & 0.05  & 0.06  & 0.05  & 0.04  & 0.10  & 0.90 \\
    \midrule
    Total to & 0.92  & 1.03  & 1.02  & 0.96  & 0.76  & 0.98  & 0.86  & 1.08  & 1.04  & 0.84  & 0.88  & 0.96  & 0.79  & 0.92  & 0.97  & 0.60  & 0.86  &  \\
\cmidrule{1-18}    \end{tabular}%
}
  \label{tab:pairwise17coinsreturn}%
\end{table}%

\begin{table}[H]
  \centering
  \caption{Pairwise volatility connectedness for sample 3}
      \resizebox{.99\textwidth}{!}{
    \begin{tabular}{lrrrrrrrrrrrrrrrrrr}
    \multicolumn{19}{c}{Period 19/07/2019 - 18/07/2020. Total connectedness: 0.87} \\
          & \multicolumn{1}{l}{ADA} & \multicolumn{1}{l}{BCH} & \multicolumn{1}{l}{BNB} & \multicolumn{1}{l}{BTC} & \multicolumn{1}{l}{DASH} & \multicolumn{1}{l}{EOS} & \multicolumn{1}{l}{ETC} & \multicolumn{1}{l}{ETH} & \multicolumn{1}{l}{LTC} & \multicolumn{1}{l}{MIOTA} & \multicolumn{1}{l}{NEO} & \multicolumn{1}{l}{TRX} & \multicolumn{1}{l}{XLM} & \multicolumn{1}{l}{XMR} & \multicolumn{1}{l}{XRP} & \multicolumn{1}{l}{XTZ} & \multicolumn{1}{l}{ZEC} & \multicolumn{1}{l}{Total from} \\
\cmidrule{2-19}    ADA   & 0.13  & 0.06  & 0.06  & 0.05  & 0.05  & 0.06  & 0.05  & 0.07  & 0.07  & 0.06  & 0.06  & 0.06  & 0.05  & 0.04  & 0.06  & 0.03  & 0.05  & 0.87 \\
    BCH   & 0.05  & 0.12  & 0.05  & 0.05  & 0.06  & 0.07  & 0.07  & 0.07  & 0.08  & 0.04  & 0.05  & 0.05  & 0.03  & 0.04  & 0.06  & 0.04  & 0.05  & 0.88 \\
    BNB   & 0.05  & 0.06  & 0.13  & 0.06  & 0.05  & 0.06  & 0.05  & 0.07  & 0.07  & 0.05  & 0.06  & 0.06  & 0.03  & 0.05  & 0.06  & 0.04  & 0.05  & 0.87 \\
    BTC   & 0.05  & 0.07  & 0.06  & 0.13  & 0.05  & 0.06  & 0.04  & 0.07  & 0.08  & 0.05  & 0.05  & 0.05  & 0.03  & 0.06  & 0.06  & 0.03  & 0.04  & 0.87 \\
    DASH  & 0.05  & 0.08  & 0.06  & 0.05  & 0.13  & 0.07  & 0.06  & 0.06  & 0.07  & 0.05  & 0.05  & 0.04  & 0.04  & 0.05  & 0.05  & 0.03  & 0.08  & 0.87 \\
    EOS   & 0.05  & 0.07  & 0.05  & 0.05  & 0.06  & 0.11  & 0.06  & 0.07  & 0.08  & 0.05  & 0.05  & 0.06  & 0.03  & 0.05  & 0.06  & 0.03  & 0.05  & 0.89 \\
    ETC   & 0.05  & 0.08  & 0.05  & 0.04  & 0.06  & 0.07  & 0.13  & 0.06  & 0.08  & 0.05  & 0.04  & 0.05  & 0.04  & 0.04  & 0.06  & 0.03  & 0.06  & 0.87 \\
    ETH   & 0.05  & 0.07  & 0.07  & 0.06  & 0.05  & 0.07  & 0.05  & 0.11  & 0.08  & 0.06  & 0.06  & 0.06  & 0.03  & 0.04  & 0.06  & 0.03  & 0.05  & 0.89 \\
    LTC   & 0.05  & 0.07  & 0.06  & 0.06  & 0.05  & 0.07  & 0.06  & 0.07  & 0.12  & 0.05  & 0.06  & 0.06  & 0.04  & 0.05  & 0.06  & 0.04  & 0.05  & 0.88 \\
    MIOTA & 0.06  & 0.06  & 0.06  & 0.05  & 0.05  & 0.06  & 0.05  & 0.07  & 0.07  & 0.13  & 0.06  & 0.05  & 0.04  & 0.04  & 0.06  & 0.04  & 0.06  & 0.87 \\
    NEO   & 0.06  & 0.06  & 0.06  & 0.05  & 0.05  & 0.06  & 0.04  & 0.07  & 0.07  & 0.05  & 0.14  & 0.06  & 0.04  & 0.04  & 0.05  & 0.04  & 0.06  & 0.86 \\
    TRX   & 0.06  & 0.06  & 0.07  & 0.04  & 0.04  & 0.06  & 0.05  & 0.06  & 0.07  & 0.06  & 0.06  & 0.13  & 0.04  & 0.04  & 0.07  & 0.03  & 0.05  & 0.87 \\
    XLM   & 0.06  & 0.06  & 0.05  & 0.04  & 0.06  & 0.05  & 0.05  & 0.06  & 0.06  & 0.06  & 0.05  & 0.06  & 0.16  & 0.04  & 0.06  & 0.03  & 0.06  & 0.84 \\
    XMR   & 0.05  & 0.06  & 0.06  & 0.06  & 0.05  & 0.06  & 0.05  & 0.06  & 0.08  & 0.05  & 0.06  & 0.05  & 0.04  & 0.13  & 0.07  & 0.03  & 0.05  & 0.87 \\
    XRP   & 0.06  & 0.06  & 0.06  & 0.05  & 0.05  & 0.06  & 0.05  & 0.06  & 0.07  & 0.06  & 0.05  & 0.07  & 0.05  & 0.05  & 0.13  & 0.03  & 0.05  & 0.87 \\
    XTZ   & 0.04  & 0.06  & 0.06  & 0.04  & 0.04  & 0.06  & 0.04  & 0.06  & 0.07  & 0.05  & 0.06  & 0.05  & 0.04  & 0.04  & 0.05  & 0.19  & 0.05  & 0.81 \\
    ZEC   & 0.05  & 0.07  & 0.05  & 0.05  & 0.08  & 0.06  & 0.05  & 0.06  & 0.07  & 0.05  & 0.06  & 0.05  & 0.04  & 0.04  & 0.05  & 0.04  & 0.14  & 0.86 \\
\cmidrule{1-18}    Total to & 0.85  & 1.05  & 0.92  & 0.80  & 0.86  & 1.02  & 0.80  & 1.05  & 1.16  & 0.84  & 0.88  & 0.86  & 0.61  & 0.71  & 0.94  & 0.54  & 0.86  &  \\
\cmidrule{1-18}    
\end{tabular}%
}
  \label{tab:pairwise17coinsvolatility}%
\end{table}%

\bibliographystyle{apalike}
\bibliography{varlsm}

\appendix

\end{document}